%% file: r6-ftp.tex
\documentstyle[11pt,amssymb,epic,eepic,epsf]{article}

\def\cE{{\cal E}}
\def\cI{{\cal I}}
\def\cL{{\cal L}}
\def\cM{{\cal M}}
\def\cQ{{\cal Q}}
\def\cR{{\cal R}}
\def\cS{{\cal S}}
\def\cT{{\cal T}}

\newcommand{\Q}{\mbox{$\cQ$}}

\renewcommand{\L}          
  {{\ifmmode{\cL}\else{$\cL$}\fi}}
\newcommand{\LL}[1]       
  {{\ifmmode{\L({\rm #1})}\else{$\L({\rm #1})$}\fi}}
\newcommand{\Lw}[1]
  {{\ifmmode{\L_w({\rm #1})}\else{$\L_w({\rm #1})$}\fi}}
\newcommand{\LQ}{\LL{\cQ}}
\newcommand{\LwQ}{\Lw{\cQ}}
\newcommand{\cc}[1]{c({\rm #1})}
\newcommand{\ccS}{c(\cS)}
\newcommand{\ccQ}{c(\cQ)}
\newcommand{\IS}[1]{{\cI\cS}({\rm #1})}
\newcommand{\ISQ}{\IS{\cQ}}
\newcommand{\MT}[1]{{\cM\cT}({\rm #1})}
\newcommand{\MTQ}{\MT{\cQ}}
\newcommand{\Fp}
  {{\ifmmode{F_p}\else{$F_p$}\fi}}
\newcommand{\FP}[1]
  {{\ifmmode{F_p({\rm #1})}\else{$F_p({\rm #1})$}\fi}}
\newcommand{\FpS}
  {{\ifmmode{F_p(\cS)}\else{$F_p(\cS)$}\fi}}
\newcommand{\FpQ}{\FP{\cQ}}
\newcommand{\crash}{{\it crash}}

\newcommand{\wQ}{{w^{\cQ}}}
\newcommand{\wS}{{w^{\cS}}}
\newcommand{\wRi}{{w^{\cR_i}}}
\newcommand{\lwS}{l_\wS}
\newcommand{\lwRi}{l_\wRi}
\newcommand{\lwQ}{l_\wQ}

\newcommand{\Mgrid}{\mbox{\rm M-Grid}}
\newcommand{\Mgridb}[1]{\mbox{{\rm M-Grid}$(#1)$}}
\newcommand{\Thresh}{\mbox{\rm Thresh}}
\newcommand{\FPP}{\mbox{\rm FPP}}
\newcommand{\bFPP}{\mbox{\rm boostFPP}}
\newcommand{\bFPPqb}
    {{\ifmmode{\bFPP\mbox{$(q,b)$}}\else{\bFPP$(q,b)$}\fi}}
\newcommand{\RT}{\mbox{\rm RT}}
\newcommand{\RTlk}
    {{\ifmmode{\RT\mbox{$(k,\ell)$}}\else{\RT$(k,\ell)$}\fi}}
\newcommand{\lofk}{\mbox{$\ell${\rm -of-}$k$}}
\newcommand{\ofiv}{\mbox{\rm 3-of-4}}
\newcommand{\klone}{k-\ell+1}
\newcommand{\crashed}{{\it \#crashed}}
\newcommand{\Mpath}{\mbox{\rm M-Path}}
\newcommand{\Mpathb}
    {{\ifmmode{\Mpath\mbox{$(b)$}}\else{\Mpath$(b)$}\fi}}
\newcommand{\sqtwo}{{\ifmmode{\sqrt{2b+1}}\else{$\sqrt{2b+1}$}\fi}}

\newcommand{\sqn}{{\ifmmode{\sqrt{n}}\else{$\sqrt{n}$}\fi}}
\renewcommand{\Pr}{{\mathbb{P}}}       
\newcommand{\ZZ}{{\mathbb{Z}}}
\newcommand{\RR}{{\mathbb{R}}}
\newcommand{\emp}{\varnothing}    

\newcommand{\tends}[1]
   {\mathrel{\mathop{\longrightarrow}\limits_{#1\to\infty}}}

\newtheorem{theorem}{Theorem}[section]
\newtheorem{proposition}[theorem]{Proposition}
\newtheorem{definition}[theorem]{Definition}

\newtheorem{claim}[theorem]{Claim}
\newtheorem{lemma}[theorem]{Lemma}

\newtheorem{corollary}[theorem]{Corollary}

\def\Proof{\par\noindent{\em Proof:~}}
\def\proof{\Proof}

\def\Remark{\noindent{\bf Remark:~}}
\newcommand{\Remarks}{\noindent{\bf Remarks: }}

\def\qed{\Box}
\def\QED{\quad$\qed$\lower 7pt\null\par}
\def\inQED{\quad\quad\Box}

\newcommand{\filefig}[4]{
  \begin{figure}[t]
    \begin{center}
      \setlength{\epsfxsize}{#4}
      \leavevmode
      \epsfbox {#2}
      \caption{\protect {#1}}
      \label {#3}
    \end{center}
  \end{figure}}

\title{The Load and Availability of Byzantine Quorum Systems\thanks{
{\bf Preprint of paper to appear in SIAM Journal of Computing.}}
}
\author{
\begin{tabular}{ccc}
Dahlia Malkhi & Michael K.\ Reiter & Avishai Wool \\ \\
AT\&T Labs -- Research &
\multicolumn{2}{c}{Bell Laboratories} \\
Florham Park, New Jersey, USA &
\multicolumn{2}{c}{Murray Hill, New Jersey, USA} \\
{\sf dalia@research.att.com} & \multicolumn{2}{c}{\sf
\{reiter,yash\}@research.bell-labs.com}
\end{tabular}
}

\begin{document}
\maketitle

\begin{abstract}
Replicated services accessed via {\em quorums} enable each access to
be performed at only a subset (quorum) of the servers, and achieve
consistency across accesses by requiring any two quorums to intersect.
Recently, $b$-masking quorum systems, whose intersections contain at
least $2b+1$ servers, have been proposed to construct replicated
services tolerant of $b$ arbitrary (Byzantine) server failures.  In
this paper we consider a hybrid fault model allowing benign failures
in addition to the Byzantine ones. We present four novel constructions
for $b$-masking quorum systems in this model, each of which has
optimal {\em load} (the probability of access of the busiest server)
or optimal availability (probability of some quorum surviving
failures). To show optimality we also prove lower bounds on the load
and availability of any $b$-masking quorum system in this model.
\end{abstract}

 
\section{Introduction}

Quorum systems are well known tools for increasing the efficiency of
replicated services, as well as their availability when servers may
fail benignly.  A quorum system is a set of subsets (quorums) of
servers, every pair of which intersect.  Quorum systems enable each
client operation to be performed only at a quorum of the servers,
while the intersection property makes it possible to preserve
consistency among operations at the service.

Quorum systems work well for environments where servers may fail
benignly.  However, when servers may suffer arbitrary (Byzantine)
failures, the intersection property does not suffice for maintaining
consistency; two quorums may intersect in a subset containing {\em
faulty} servers only, who may deviate arbitrarily and undetectably
from their assigned protocol.  Malkhi and Reiter thus introduced {\em
masking quorums systems}~\cite{mr98a}, in which each pair of quorums
intersects in sufficiently many servers to mask out the behavior of
faulty servers.  More precisely, a {\em $b$-masking quorum system} is
one in which any two quorums intersect in $2b+1$ servers, which
suffices to ensure consistency in the system if at most $b$ servers
suffer Byzantine failures.

In this paper we develop four new constructions for $b$-masking quorum
systems.  For the first time in this context, we distinguish between
masking Byzantine faults and surviving a possibly larger number of
benign faults. Our systems remain available in the face of any $f$
crashes, where $f$ may be significantly larger than $b$ (such a system
is called $f$-resilient). In addition, our constructions demonstrate
optimality (ignoring constants) in two widely accepted measures of
quorum systems, namely {\em load} and {\em crash probability}.  The
load (\L), a measure of best-case performance of the quorum system, is
the probability with which the busiest server is accessed under the
best possible strategy for accessing quorums.  The crash probability
($F_p$) is the probability, assuming that each server crashes with
independent probability $p$, that all quorums in the system will
contain at least one crashed server (and thus will be
unavailable). The crash probability is an even more refined measure of
availability than $f$, as a good system will tolerate many failure
configurations with more than $f$ crashes.  Three of our systems are
the first systems to demonstrate optimal load for $b$-masking quorum
systems, and two of our systems each demonstrate optimal crash
probability for its resilience $f$.  In proving optimality of our
constructions, we prove new lower bounds for the load and crash
probability of masking quorum systems.

The techniques for achieving our constructions are of interest in
themselves.  Two of the constructions are achieved using a {\em
boosting\/} technique, which can transform any regular (i.e., benign
fault-tolerant) quorum system into a masking quorum system of an
appropriately larger system.  Thus, it makes all known quorum
constructions available for Byzantine environments (of appropriate
sizes). In the analysis of one of our best systems we employ strong
results from percolation theory.

The rest of this paper is structured as follows.  We review related
work and preliminary definitions in Sections~\ref{related}
and~\ref{prelims}, respectively.  In Section~\ref{blocks} we prove
bounds on the load and crash probability for $b$-masking quorum
systems and introduce quorum composition.  In
Sections~\ref{simple}--\ref{mpaths} we describe our new constructions.
We discuss our results in Section~\ref{discussion}.

\section{Related work} \label{related} 

Our work borrows from extensive prior work in benignly fault-tolerant
quorum systems
(e.g.,~\cite{gif79,tho79,mae85,gb85,her86,bg87,et89,ae91,caa92,nw98,pw97a}).
The notion of availability we use here (crash probability) is well
known in reliability theory \cite{bp75} and has been applied
extensively in the analysis of quorum systems (cf.\
\cite{bg87,pw95d,pw97b} and the references therein). The load of a
quorum system was first defined and analyzed in \cite{nw98}, which
proved a lower bound of $\Omega(\frac{1}{\sqrt{n}})$ on the load of
any quorum system (and, a fortiori, any masking quorum system) over
$n$ servers.  In proving load-optimality of our constructions, we
generalize this lower bound to $\Omega(\sqrt{\frac{b}{n}})$ for
$b$-masking quorum systems.

Grids, which form the basis for our M-Grid construction, were proposed
in~\cite{mae85,caa92,krs93,mr98a}.  The technique of quorum composition,
which we use in our $\RT$ and \bFPP\ constructions, has been studied
in \cite{mp92a,nm92,nei92} under various names such as ``coterie
join'' and ``recursive majority''.  Our \Mpath\ construction
generalizes the system of \cite{wb92}, coupled with the analysis of
the Paths construction of~\cite{nw98}, and the recent system
of~\cite{baz96}.

Several constructions of masking quorum systems were given
in~\cite{mr98a} for a variety of failure models.  For the model we
consider here---i.e., any $b$ servers may experience Byzantine
failures---that work gave two constructions.  We compare those
constructions to ours in Section~\ref{discussion}.

Hybrid failure models have been considered in other works
(e.g.,~\cite{gp92,lr93,lr94,rb94}).

\section{Preliminaries} \label{prelims} 

In this section we introduce notation and definitions used in the
remainder of the paper.  Much of the notation introduced in this
section is summarized in Table~\ref{table:notation} for quick
reference.

\begin{table}[t]
\begin{center}
\begin{tabular}{|c|l|}
\hline
$b$    & Maximum number of Byzantine server failures. \\
$\ccQ$ & Size of the smallest quorum in $\cQ$. \\
$f$    & Resilience (Definition~\ref{defResil}). \\
$\FpQ$ & Crash probability (Definition~\ref{DefFp}). \\
$\ISQ$ & Size of smallest intersection between any two quorums in $\cQ$.\\
$\LQ$  & Load of $\cQ$ (Definition~\ref{defLoad}). \\
$\MTQ$ & Size of a smallest transversal of $\cQ$
         (Definition~\ref{defMinTrans}). \\
$n$    & Number of servers (i.e., $|U| = n$). \\
$p$    & Independent probability that each server crashes. \\
$\cQ$  & A quorum system (Definition~\ref{defQuorumSys}). \\
$U$    & Universe of servers. \\ \hline
\end{tabular}
\end{center}
\caption{The notation used in this paper.}
\label{table:notation}
\end{table}

We assume a {\em universe} $U$ of servers, $|U| = n$, over which our
quorum systems will be constructed.  Servers that obey their
specifications are {\em correct}.  A {\em faulty} server, however, may
deviate from its specification arbitrarily.  We assume that up to $b$
servers may fail arbitrarily and that $4b < n$, since this is
necessary for a $b$-masking quorum system to exist~\cite{mr98a}.
Beginning in Section~\ref{availability}, we will also distinguish
benign (crash) failures as a particular failure of interest, and in
general there may be more than $b$ such failures.

\subsection{Quorum systems}

\begin{definition} \label{defQuorumSys}
A {\em quorum system\/} $\cQ \subseteq 2^U$ is a collection of subsets
of $U$, each pair of which intersect. Each $Q \in \cQ$ is called
a {\em quorum}. 
\end{definition}

We use the following notation.  The cardinality of the smallest quorum
in $\cQ$ is denoted by $\ccQ = \min\{|Q| : Q \in \cQ\}$.  The size of
the smallest intersection between any two quorums is denoted by $\ISQ
= \min\{|Q \cap R| : Q,R \in \cQ\}$.  The degree of an element
$i\in U$ in a quorum system $\cQ$ is the number of quorums that
contain $i$: $\deg(i) = | \{ Q\in\cQ : i\in Q \} |$.
\begin{definition}  \label{defFair}
A quorum system $\cQ$ is {\em $(s,d)$-fair} 
if $|Q|=s$ for all $Q\in\cQ$ and $\deg(i) =d$ for all $i\in U$. 
$\cQ$ is called {\em fair} if it is $(s,d)$-fair for some $s$ and $d$.
\end{definition}
\begin{definition} \label{defMinTrans}
A set $T$ is a {\em transversal} of a quorum system $\cQ$ if 
$T\cap Q\ne\emp$  for every 
$Q\in\cQ$. The cardinality of the smallest
transversal is denoted by $\MTQ = \min\{|T| : T {\rm\ is\ a\
  transversal\ of\ } \cQ\}$. 
\end{definition}

Regular quorum systems, with $\ISQ=1$, are insufficient to guarantee
consistency in case of Byzantine failures.  Malkhi and
Reiter~\cite{mr98a} defined several varieties of quorum systems for
Byzantine environments, which are suitable for different types of
services. In this paper we focus on {\em masking quorum systems}.

\begin{definition} \label{defResil} 
  {\rm \cite{mr98a}} The {\em resilience} $f$ of a quorum system \Q\ is the
  largest $k$ such that for every set $K \subseteq U$, $|K| = k$,
  there exists $Q\in \Q$ such that $K \cap Q = \emp$.
\end{definition}

\Remark The resilience of any quorum system $\cQ$ is $f = \MTQ - 1$.

\begin{definition} \label{defMask} 
{\rm \cite{mr98a}}
A quorum system $\cQ$ is a {\em $b$-masking quorum system} if it is
resilient to $f \ge b$ failures, and obeys the following {\em
consistency} requirement:
\begin{equation}
 \forall Q_1, Q_2 \in \cQ: |Q_1 \cap Q_2| \ge 2b+1. \label{eq:masking}
\end{equation}
\end{definition}

\Remark Informally, if we view the service as a shared variable which
is updated and read by the clients, then the resilience requirement of
Definition~\ref{defResil} ensures that no set of $b\le f$ faulty
servers will be able to block update operations (e.g., by causing
every update transaction to abort). The consistency requirement of
Definition~\ref{defMask} ensures that read operations can mask out any
faulty behavior of up to $b$ servers.  Examples of protocols
implementing various data abstractions using $b$-masking quorum
systems can be found in~\cite{mr98a,mr98b,mr98c}.

\begin{lemma}  \label{lemMaskCond} 
  Let \Q\ be a quorum system. Then \Q\ is $b$-masking if both the
  following conditions hold:
  \begin{enumerate}
  \item $\MTQ \ge b+1$,
  \item $\ISQ \ge 2b+1$.
  \end{enumerate}
\end{lemma}
\proof
Assume that $\MTQ\ge b+1$. To see that \Q\ is resilient to $b$
failures, note that if there exists some $K$ such
that $K\cap Q \ne \emp$ for all $Q\in \cQ$, then $K$ is a transversal.
By the minimality we have $|K| \ge b+1$, and we are done. Condition~2
immediately implies~(\ref{eq:masking}).
\QED

\begin{corollary} \label{corMaskCond} 
  Let \Q\ be a quorum system, and let 
  $b$ $=$ $\min\{\MTQ-1$,$\frac{\ISQ-1}{2}\}$.
  Then \Q\ is $b$-masking. \QED
\end{corollary}

\subsection{Measures}

The goal of using quorum systems is to increase the availability of
replicated services and decrease their access costs.  A natural
question is how well any particular quorum system achieves these
goals, and moreover, how well it compares with other quorum systems.
Several measures will be of interest to us.

\subsubsection{Load}
A measure of the inherent performance of a quorum system is its {\em
  load}.  Naor and Wool define the load of a quorum system as the
frequency of accessing the busiest server using the best possible
strategy~\cite{nw98}. More precisely, given a quorum system $\cQ$, an
{\em access strategy} $w$ is a probability distribution on the
elements of $\cQ$; i.e., $\sum_{Q \in \cQ} w(Q) = 1$.  The value
$w(Q) \geq 0$ is the frequency of choosing quorum $Q$ when the
service is accessed. The load is then defined as follows:

\begin{definition} \label{defLoad}
Let a strategy $w$ be given for a quorum system 
$\cQ = \{Q_1,\ldots,Q_m\}$ over a universe $U$.  
For an element $u \in U$,
the load induced by $w$ on $u$ is $l_w(u) = \sum_{Q_i \ni u} w(Q_i)$.
The load induced by a strategy $w$ on a quorum system $\cQ$ is
$
\LwQ = \max_{u \in U}\{l_w(u)\}.
$
The {\em system load} on a quorum system $\cQ$ is
$
\LQ = \min_{w}\{\LwQ\},
$
where the minimum is taken over all strategies.
\end{definition}

We reiterate that the load is a best case definition.  The load of the
quorum system will be achieved only if an optimal access strategy is
used, and only in the case that no failures occur.  A strength of this
definition is that the load is a property of a quorum system, and not
of the protocol using it.  Examples of load calculations can be found
in~\cite{woo96}. As an aside, we note that not every quorum system can
have a strategy that induces the same load on each server. In
\cite{hmp97} it is shown that for some quorum systems it is impossible
to perfectly balance the load.

Recall that $\ccQ$ denotes the cardinality of the smallest quorum in $\cQ$. 
The following result will be useful to us in the sequel
(recall Definition~\ref{defFair}).
\begin{proposition} \label{pFair} 
{\rm \cite{nw98}}
Let $\cQ$ be a fair quorum system. Then $\LQ = \ccQ/n$.
\end{proposition}

\subsubsection{Availability}
\label{availability}

By definition a $b$-masking quorum system can mask up to $b$ arbitrary
(Byzantine) failures. However, such a system may be resilient to more
{\em benign} failures.  By benign failures we mean any failures that
render a server unresponsive, which we refer to as {\em crashes} to
distinguish them from Byzantine failures.

The resilience $f$ of a quorum system provides one measure of how many
crash failures a quorum system is {\em guaranteed} to survive, and indeed
this measure has been used in the past to differentiate among quorum
systems~\cite{bg86}.  However, it is possible that an $f$-resilient
quorum system, though vulnerable to a few failure configurations of
$f+1$ failures, can survive many configurations of more than $f$
failures.  One way to measure this property of a quorum system is to
assume that each server crashes independently with probability $p$ and
then to determine the probability \Fp\ that some quorum survives with
no faulty members.  This is known as {\em crash probability} and is
formally defined as follows:
\begin{definition} \label{DefFp}
Assume that each server in the system cra\-shes independently with
probability $p$.
For every quorum $Q\in \cQ$ let $\cE_Q$ be the event that $Q$ is {\em hit},
i.e., at least one element $i\in Q$ has crashed.
Let $\crash(\cQ)$ be the event that all the quorums $Q\in \cQ$ were hit,
i.e., $\crash(\cQ) = \bigwedge_{Q\in\cQ} \cE_Q$.
Then the system crash probability is $\FpQ = \Pr(\crash(\cQ))$.
\end{definition}

We would like \Fp\ to be as small as possible.  A desirable asymptotic
behavior of \Fp\ is that $\Fp \to 0$ when $n\to\infty$ for {\em all\/}
$p<1/2$, and such an \Fp\ is called Condorcet (after the Condorcet
Jury Theorem \cite{condor}).

\section{Building blocks}   \label{blocks} 

In this section, we prove several theorems which will be our basic tools
in the sequel. First we prove lower bounds on the load and
availability of $b$-masking
quorum systems, against which we measure all our new
constructions. Then we prove the properties of a quorum composition
technique, which we later use extensively.

\subsection{The load and availability of masking quorum systems}

We begin by establishing a lower bound on the load of $b$-masking
quorum systems, thus tightening the lower bound on general quorum
systems~\cite{nw98} as presented in~\cite{mr98a}.

\begin{theorem} \label{thmLoadBound} 
Let \Q\ be a $b$-masking quorum system. Then 
$
\LQ \ge \max \{ \frac{2b+1}{\ccQ}, \frac{\ccQ}{n} \}.
$
\end{theorem}
\proof
Let $w$ be any strategy for the quorum system $\Q$, and fix $Q_1
\in \Q$ such that $|Q_1| = c(\Q)$.  Summing the loads
induced by $w$ on all the elements of $Q_1$,
and using the fact that any two quorums have at least $2b+1$ elements
in common, we obtain:
\begin{eqnarray*}
   \sum_{u \in Q_1} l_w(u) &=& \sum_{u \in Q_1} \sum_{Q_i \ni u} w(Q_i)
       ~=~ \sum_{Q_i} \sum_{u \in (Q_1 \cap Q_i)} w(Q_i) \\
        &\geq& \sum_{Q_i} (2b+1)w(Q_i) 
        ~=~  2b+1.
\end{eqnarray*}
Therefore, there exists some element in $Q_1$ that suffers a load of
at least $\frac{2b+1}{|Q_1|}$.

Similarly, summing the total load induced by $w$ on all of the
elements of the universe, and using the minimality of $\ccQ$, we get:

\begin{eqnarray*}
     \sum_{u \in U} l_w(u) &=& \sum_{u \in U} \sum_{Q_i \ni u} w(Q_i)
        ~=~ \sum_{Q_i} |Q_i| w(Q_i) \\
        &\geq& \sum_{Q_i} \ccQ w(Q_i) ~=~ c(\Q).
\end{eqnarray*}
Therefore, there exists some element in $U$ that suffers a load of at
least $\frac{\ccQ}{n}$.
\QED

\begin{corollary}   \label{corLowBound} 
Let \Q\ be a $b$-masking quorum system. Then 
$
 \LQ \ge \sqrt{\frac{2b+1}{n}},
$
and equality holds if $\ccQ=\sqrt{(2b+1)n}$.\footnote{To avoid
repetitive notation, we omit floor and ceiling brackets from
expressions for integral quantities.} \QED
\end{corollary}

\Remark Corollary~\ref{corLowBound} shows that the threshold
construction of \cite{mr98a} in fact has optimal load when
$b=\Omega(n)$. E.g., when $b\approx n/4$ the obtained load is $\approx
0.75$, but for such systems we can only hope for a constant load of
$\approx 1/\sqrt{2}=0.707$. However the load of the threshold
construction is always $\ge 1/2$, which is far from optimal for
smaller values of~$b$.

On the other hand, the grid-based construction of \cite{mr98a} does not
have optimal load.  It has quorums of size $O(b\sqn)$ and load of
roughly $2b/\sqn$.  In the sequel we show systems which significantly
improve this: some of our new constructions have quorums of size
$O(\sqrt{bn})$ and optimal load.

\bigskip

Our next propositions show lower bounds on the crash probability \Fp\
in terms of~$\MTQ$ and~$b$.
 
\begin{proposition}                    \label{pMT} 
Let \Q\ be a quorum system. Then $\FpQ \ge p^{\MTQ}=p^{f+1}$ for any
$p\in[0,1]$.
\end{proposition}
\proof
Consider a minimal transversal $T$ with $|T|=\MTQ$. If all the elements of
$T$ crash then every quorum contains a crashed element, so
$\FpQ \ge p^{\MTQ}$. 
\QED

\begin{proposition}  \label{pC2F}
Let \Q\ be a $b$-masking quorum system. Then $\FpQ \ge p^{\ccQ-2b}$ for
any $p\in[0,1]$.
\end{proposition}
\proof
Let $Q\in\cQ$ be a minimal quorum with $|Q|=\ccQ$, and consider
$Z\subset Q$, $|Z|=2b$. Since \Q\ is $b$-masking then $|R\cap Q|\ge 2b+1$
for any $R\in\cQ$, and so $|(Q\setminus Z)\cap R|\ge 1$ and 
$Q\setminus Z$ is a transversal. Therefore $\MTQ\le \ccQ-2b$, which we
plug into Proposition~\ref{pMT}.
\QED

\noindent The next proposition is less general than
Proposition~\ref{pC2F}, however it is applicable for most of our
constructions and it gives a much tighter bound.

\begin{proposition}                    \label{pF} 
Let \Q\ be a $b$-masking quorum system such that $\MTQ \le (\ISQ+1)/2$. 
Then $\FpQ \ge p^{b+1}$ for any $p\in[0,1]$.
\end{proposition}
\proof
If $\MTQ \le (\ISQ+1)/2$ then from
Corollary~\ref{corMaskCond} we have that $b+1=\MTQ$, which again we
plug into Proposition~\ref{pMT}.
\QED



\subsection{Quorum system composition}

Quorum system composition is a well known technique for building new
systems out of existing components.  We compose a quorum system $\cS$
over another system $\cR$ by replacing each element of $\cS$ with a
distinct copy of $\cR$.  In other words, when element $i$ is used in a
quorum $S\in\cS$ we replace it with a complete quorum from the $i$'th
copy of $\cR$.  Using the terminology of reliability theory, the
system $\cS\circ\cR$ has a {\em modular decomposition} where each
module is a copy of $\cR$. Formally:

\begin{definition} \label{defCompose} 
  Let $\cS$ and $\cR$ be two quorum systems, over universes of sizes
  $n_S$ and $n_R$, respectively. Let $\cR_1,\ldots,\cR_{n_S}$ be $n_S$
  copies of $\cR$ over disjoint universes. Then the {\em composition}
  of $\cS$ over $\cR$ is
  \[
  \cS\circ\cR = \left\{ \bigcup R_i : S\in\cS, 
     R_i\in\cR_i {\rm\ for\ all\ }i\in S\right\}.
  \]
\end{definition}

\noindent
The next theorem summarizes the properties of quorum composition.

\begin{theorem} \label{thmComposition} 
  Let $\cS$ and $\cR$ be two quorum systems, and let
  $\cQ=\cS\circ\cR$. Then
  \begin{itemize}
  \item The universe size is $n_Q=n_S n_R$.
  \item The minimal quorum size is $\ccQ=\ccS \cc{\cR}$.
  \item The minimal intersection size is $\ISQ=\IS{\cS} \IS{\cR}$.
  \item The minimal transversal size is $\MTQ=\MT{\cS} \MT{\cR}$.
  \item Denote the crash probability functions of $\cS$ and $\cR$ by
    $s(p) = \FpS$ and $r(p) = \FP{\cR}$. Then $\FP{\cQ} = s(r(p))$.
  \item The load is $\LQ=\LL{\cS} \LL{\cR}$.
  \end{itemize}
\end{theorem}
\proof 
The behavior of the combinatorial parameters $n_Q$, $\ccQ$, $\ISQ$ and
$\MTQ$ is obvious. The behavior of \FpQ\ is standard in reliability
theory (cf. \cite{bp75}).  As for the load, consider the following
strategy: pick a quorum $S\in\cS$ using the optimal strategy for
$\cS$. Then for each element $i\in S$, pick a quorum $R_i\in\cR_i$
using the optimal strategy for (the $i$'th copy of) $\cR$. Clearly
this strategy induces a load of $\LL{\cS} \LL{\cR}$, and hence $\LQ
\le \LL{\cS}\LL{\cR}$.

We now show the inequality in the opposite direction.  Enumerate the
elements of $\cQ$ by denoting the $j$'th element in $\cR_i$ by
$u_{ij}$, let $Q(S) = \left\{ \bigcup R_i : R_i\in\cR_i {\rm\ for\
all\ }i\in S\right\}$ be the set of all quorums that are based on some
$S\in\cS$, and let $\wQ$ be an access strategy on $\cQ$.  Then $\wQ$
induces a strategy $\wS$ on $\cS$ defined by

\begin{equation} \label{e43}
 \wS(S) = \sum_{Q\in Q(S)} \wQ(Q). 
\end{equation}
The load on an element $i\in \cS$ (i.e., the frequency of accessing
the quorum system $\cR_i$) is then 
$\lwS(i) = \sum_{S\ni i, S\in\cS}\wS(S)$. 
Similarly, $\wQ$ induces a strategy on each copy $\cR_i$
defined by
\begin{equation} \label{e44}
  \wRi(R) = \bigg(\sum_{Q\supseteq R} \wQ(Q)\bigg) \bigg/ \lwS(i).
\end{equation}
This $\wRi$ is well defined when $\lwS(i) > 0$. It is easy to verify 
that $\wS$ and $\wRi$ are indeed strategies, i.e., that the
probabilities add up to 1. 

\begin{claim}
  Let $\lwQ(u_{ij})$ be the load induced by $\wQ$ 
  on an element $u_{ij} \in \cR_i$, and let $\lwRi(u_{ij})$ be the
  load induced on it by $\wRi$. 
  Then $\lwQ(u_{ij}) = \lwS(i)\cdot\lwRi(u_{ij})$.
\end{claim}
{\it Proof of Claim:} 
Using (\ref{e43}) and (\ref{e44}) we have that
\begin{eqnarray*}
 \lwS(i)\cdot\lwRi(u_{ij}) &=& \lwS(i)\sum_{R\ni u_{ij}}\wRi(R) ~=~
 \lwS(i)\sum_{R\ni u_{ij}}
           \bigg(\sum_{Q\supseteq R} \wQ(Q)\bigg)\bigg/\lwS(i) \\
 &=& \sum_{R\ni u_{ij}}\sum_{Q\supseteq R} \wQ(Q) 
  ~=~ \sum_{Q\ni u_{ij}}\wQ(Q) ~=~ \lwQ(u_{ij}). \inQED
\end{eqnarray*}

To complete the proof of Theorem~\ref{thmComposition}, assume that
$\wQ$ is an optimal strategy for $\cQ$. Consider the copy $\cR_i$ for
which $\lwS(i)$ is maximal, i.e., $\L_{\wS}(\cS) = \lwS(i)$, and let
$u_{ij}$ be the maximally-loaded element in this $\cR_i$.  Clearly
$\lwS(i)>0$ so $\wRi$ is well defined for this $i$.  Note that
we do not require $u_{ij}$ to be the maximally-loaded element in all of
$\cQ$. Using the claim and the minimality of $\LL{\cS}$ and
$\LL{\cR}$ we obtain that
\begin{eqnarray*}
  \LQ &=& \L_{\wQ}(\cQ) ~\ge~ \lwQ(u_{ij}) ~=~
                \lwS(i)\cdot\lwRi(u_{ij}) \\
      &=& \L_{\wS}(\cS)\cdot\L_{\wRi}(\cR) ~\ge~ \LL{\cS}\LL{\cR}.
\end{eqnarray*}
By combining this inequality with the upper bound we had before we
conclude that $\LQ = \LL{\cS}\LL{\cR}$.
\QED

The multiplicative behavior of the combinatorial parameters in
composing quorum systems provides a powerful tool for ``boosting''
existing constructions into larger systems with possibly improved
characteristics. Below, we use quorum composition in two cases, and
demonstrate that this technique yields improved constructions over
their basic building blocks, for appropriately larger system sizes.  In
particular, in Section~\ref{secBFPP} we show a composition
that allows us to transform any regular quorum construction into a
(larger) $b$-masking quorum system.

\section{Simple systems}   \label{simple} 

In this section we show two types of constructions, the multi-grid
(denoted \Mgrid) and the recursive threshold (\RT).  These systems
significantly improve upon the original constructions of \cite{mr98a},
however both are still suboptimal in some parameter: \Mgrid\ has
optimal load but can mask only up to $b=O(\sqn)$ failures and has
poor crash probability; and \RT\ can mask up to $b=O(n)$ failures
and has near optimal crash probability, but has suboptimal load.

In Sections~\ref{secBFPP} and \ref{mpaths} we present systems which
are superior to the
\Mgrid\ and \RT. Nonetheless, we feel that the simplicity of the
\Mgrid\ and \RT\ systems, and the fact that they are suitable for
very small universe sizes, are what makes them appealing.

\subsection{The multi-grid system}

\begin{figure}
\centerline{\input{grid.tex}}
\caption{The multi-grid construction, $n = 7 \times 7, b = 3$, with
one quorum shaded.}
\label{grid}
\end{figure}
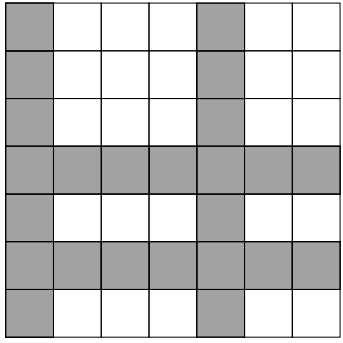

We begin with the \Mgrid\ system, which achieves an optimal load among
$b$-masking quorum systems, where $b \le (\sqn-1)/2$.  The idea of the
construction is as follows.  Arrange the elements in a
$\sqrt{n}\times \sqrt{n}$ grid. A quorum in a multi-grid consists of
any choice of $\sqrt{b+1}$ rows and $\sqrt{b+1}$ columns, as shown in
Figure~\ref{grid}. Formally, denote the rows and columns of the grid
by $R_i$ and $C_i$, respectively, where $1 \le i \le \sqrt{n}$. Then,
the quorum system is
\[
 \Mgridb{b} =
   \left\{ \bigcup_{j \in J} C_j \cup \bigcup_{i \in I} R_i : 
        J, I \subseteq \{ 1 \ldots \sqrt{n} \}, |J| = |I| = \sqrt{b+1}
        \right\}. 
\]
\begin{proposition}
  The multi-grid \Mgridb{b} is a $b$-masking quorum system for
$b\le(\sqn-1)/2$.
\end{proposition}
\proof
Consider two quorums $R,S\in\Mgridb{b}$. If they have either a row or a
column in common, then $|R\cap S|\ge\sqn\ge 2b+1$ and we are done. Otherwise
the intersection of $S$'s columns with $R$'s rows is disjoint from 
the intersection of $R$'s columns with $S$'s rows, so
$|R\cap S|\ge 2\sqrt{b+1}\sqrt{b+1} > 2b+1$. Therefore consistency holds.

Resilience holds since $f=\MT{\Mgridb{b}}-1=\sqn-\sqrt{b+1} \ge b$.
Therefore $\MT{\Mgridb{b}} \ge b+1$, and Lemma \ref{lemMaskCond} finishes
the proof.  
\QED

\begin{proposition}
  $\LL{\Mgridb{b}} \approx 2\sqrt{\frac{b+1}{n}}$.
\end{proposition}
\proof
Since \Mgridb{b} is fair we can use Proposition~\ref{pFair} to
get $\LL{\Mgridb{b}}=\cc{\Mgridb{b}}/n$.
\QED

\Remark The load of \Mgridb{b} is within a factor of $\sqrt{2}$ from 
the optimal load which can be achieved for $b\approx\sqn/2$.

A disadvantage of the \Mgrid\ system is its poor asymptotic crash
probability.  If crashes occur with some constant probability $p$ then
any configuration of crashes with at least one crash per row disables
the system. Therefore, as shown by \cite{kc91,woo96},
\[
  \FP{\Mgrid} \ge (1-(1-p)^\sqn)^\sqn \tends{n} 1.
\]

\subsection{Recursive threshold systems}

\begin{figure} 
\centerline{\input{lofk.tex}}
\caption{An $\RT(4,3)$ system of depth $h = 2$, with one quorum
  shaded.}
\label{figRT} 
\end{figure}
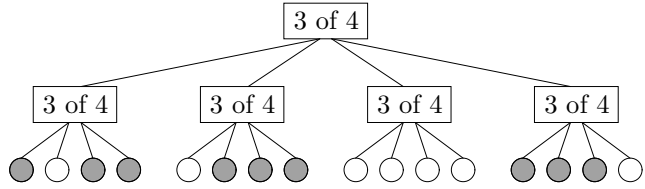

A recursive threshold system \RTlk\ of depth $h$ is built by taking a 
simple building block, which is an \lofk\ threshold system
(with $k>\ell>k/2$), 
and recursively composing it over itself to depth $h$.
In the sequel, we often omit the depth parameter $h$ when it has no
effect on the discussion.
The \RT\ systems generalize the recursive majority
constructions of \cite{mp92a}, the HQS system of
\cite{kum91} is an $\RT(3,2)$ system, and in fact the threshold system of
\cite{mr98a} can be viewed as a trivial $\RT(4b+1,3b+1)$ 
system with depth $h=1$.
As an example throughout this section we will use the $\RT(4,3)$
system, depicted in Figure~\ref{figRT}. 

\begin{proposition}
  An \RTlk\ system of depth $h$ is a fair quorum system, with $n=k^h$
  elements, quorums of size 
  $\cc{\RTlk}=\ell^h$, intersection size of
  $\IS{\RTlk}=(2\ell-k)^h$, and minimal transversals of size
  $\MT{\RTlk}=(\klone)^h$.
\end{proposition}
\proof 
The basic \lofk\ system is symmetric (and therefore fair), with
$\cc{\lofk}=\ell$, $\MT{\lofk}=\klone$, and $\IS{\lofk}=2\ell-k$.
The combinatorial parameters are computed by activating
Theorem~\ref{thmComposition} $h$ times, and the composition preserves
the fairness.  
\QED

\noindent Plugging this into Corollary~\ref{corMaskCond} we obtain
\begin{corollary}
An \RTlk\ system over a universe of size $n$ is a $b$-masking quorum
system for 
\[
b=\min\{{(n^{\log_k(2\ell-k)}-1)/2},n^{\log_k(\klone)}-1\}. \inQED
\]
\end{corollary}
\noindent In the \ofiv\ example we have $\IS{\ofiv}=\MT{\ofiv}=2$ and
$\cc{\ofiv}=3$. Therefore for the whole system (to depth $\log_4 n$)
we get $\cc{\RT(4,3)}=n^{\log_4 3}=n^{0.79}$, with
$\IS{\RT(4,3)}=\MT{\RT(4,3)}=\sqn$ and thus 
$b=(\sqn-1)/2$. Note that the basic \ofiv\ system is not even
1-masking since intersections of size~2 are too small, however
already from $h=2$ (i.e., $n=16$) we obtain a masking system.

\begin{proposition}
  The load $\LL{\RTlk} = n^{-(1-\log_k \ell)}$.
\end{proposition}
\proof
Since \RTlk\ is fair we can use Proposition~\ref{pFair} to
get $\LL{\RTlk}=\cc{\RTlk}/n$.
\QED

\Remark In general the load is suboptimal for this construction.
For instance, in the $\RT(4,3)$ system we obtain 
$\LL{\RT(4,3)}=n^{-0.21}$. However for $b=(\sqn-1)/2$ we
could hope for a load of $\sqrt{(2b+1)/n}=n^{-0.25}$.

\begin{proposition} \label{pFpLimit} 
  There exists a unique critical probability $0<p_c<1/2$ for which 
\[
   \lim_{h\to\infty}\FP{\RTlk ~\mbox{\rm of depth $h$}} = 
                \cases{0, & $p<p_c$,\cr 1, & $p>p_c$.\cr }
\]
\end{proposition}
\proof
Let $g(p)$ be the crash probability function of the \lofk\ system
and let $F(h) =$ $\FP{\RTlk {\rm\ of\ depth\ }h}$ denote
the crash probability for the \RTlk\ system of depth $h$. Then
$F(h)$ obeys the recurrence
\begin{equation}
  \label{e40}
   F(h) = \cases{g( F(h-1)), & $h\ge 1$,\cr
                 p,          & $h=0$.\cr   }  
\end{equation}
Now $g(p)$ is a reliability function, and therefore it is ``S-shaped''
(see \cite{bp75}). This implies that there exists a unique critical
probability $0<p_c<1$ for which $g(p_c) = p_c$, such that $g(p)<p$
when $p<p_c$ and $g(p)>p$ when $p>p_c$ (and \cite{pw95d} shows that
for quorum systems such as \RT\ in fact $p_c<1/2$).  Therefore if
$p<p_c$ then repeated applications of recurrence (\ref{e40}) would
decrease $F(h)$ arbitrarily close to~$0$, and when $p>p_c$ the limit
is~1.
\QED

\begin{proposition} \label{pRTFp} 
   If $p<1/{k\choose{\ell-1}}$ and $\ell<k$ then 
   $\FP{\RTlk} < \exp(-\Omega(n^{\log_k(\klone)}))$, which is
   optimal for systems with resilience $f=n^{\log_k(\klone)}$.
\end{proposition}
\proof
Let $g(p)$ and $F(h)$ be as in the proof of
Proposition~\ref{pFpLimit}.  Any configuration of at least $\klone$
crashes disables the \lofk\ system, so
\[
   g(p)=\sum_{j=\klone}^k{k\choose j}p^j(1-p)^{k-j}.
\]
By Lemma~\ref{lApp1} (see Appendix) we have that
\[
g(p) \le {k\choose {\ell-1}}p^{\klone}.
\]
Plugging this into (\ref{e40}) gives that 
\begin{eqnarray*}
 F(h) &\le& {k\choose {\ell-1}}^{1+(\klone)+\cdots+(\klone)^{h-1}}
        p^{(\klone)^h} \\
  &<& \left[{k\choose {\ell-1}}p\right]^{(\klone)^h}.
\end{eqnarray*}
If $p<1/{k\choose {\ell-1}}$ then the last expression decays to zero
with $h$, so  $\FP{\RTlk} < \exp(-\Omega(n^{\log_k(\klone)}))$.

The lower bound of
Proposition~\ref{pMT} shows that 
\[
\FP{\RTlk}\ge p^{n^{\log_k(\klone)}},
\]
so our analysis is tight.
\QED

For the $\RT(4,3)$ system a direct calculation shows that 
$g(p) = 6p^2-8p^3+3p^4$ and $p_c=0.2324$.  Therefore
Proposition~\ref{pFpLimit} guarantees that when the element crash
probability is in the range $p<0.2324$ then $\Fp\to 0$ when
$n\to\infty$. Furthermore, when $p<1/6$
then Proposition~\ref{pRTFp} shows that the decay is rapid,
with $\FP{\RT(4,3)}<(6p)^{\sqn}$, which is optimal.

\section{Boosted finite projective planes}   \label{secBFPP} 

In this section we introduce a family of $b$-masking quorum systems,
the {\em boosted finite projective planes}, which we denote by \bFPP.
A \bFPP\ system is a composition of a finite projective plane (\FPP)
over a threshold system (\Thresh).

The first component of a \bFPP\ system is a finite projective plane of
order $q$ (a good reference on \FPP's is \cite{hal86}). 
It is known that \FPP's exist for $q=p^r$ when $p$ is prime. 
Such an \FPP\ has $n_F = q^2+q+1$ elements, and quorums of size
$\cc{\FPP}=q+1$.  This is a regular quorum system, i.e., it has
intersections of size $\IS{\FPP}=1$. The minimal transversals of an \FPP\
are of size $\MT{\FPP}=q+1$ (in fact the only transversals of this size
are the quorums themselves). The load of \FPP\ was analyzed in
\cite{nw98} and shown to be 
 $\LL{\FPP}=\frac{q+1}{n_F}\approx 1/\sqrt{n_F}$, 
which is optimal for regular quorum systems.

The second component of a \bFPP\ is a \Thresh\ system, with $n_T=4b+1$
elements and a threshold of $3b+1$. This is a $b$-masking quorum
system in itself, with $\IS{\Thresh}=2b+1$, $\MT{\Thresh}=b+1$ and a load of
$\LL{\Thresh} \approx 3/4$.

\begin{proposition}
  Let $\bFPPqb = \FPP(q) \circ \Thresh(3b+1 {\rm\ of\ }4b+1)$. Then 
the composed system has $n=(4b+1)(q^2+q+1)$ elements, with quorums of size
$\cc{\bFPPqb} = (3b+1)(q+1)$, 
intersections of size $\IS{\bFPPqb}=2b+1$ and 
minimal transversals of size $\MT{\bFPPqb}=(b+1)(q+1)$.
Therefore \bFPPqb\ is a $b$-masking quorum system.
\end{proposition}
\proof
We obtain the combinatorial parameters by plugging the values of the
component systems into Theorem~\ref{thmComposition}. By
Corollary~\ref{corMaskCond} we have that the system can mask
$\min\{(b+1)(q+1)-1, b)\}=b$ failures.
\QED

\begin{proposition}
 $\LL{\bFPPqb} \approx \frac{3}{4q}$, 
 which is optimal for
 $b$-masking quorum systems with $n\approx 4bq^2$ elements.
\end{proposition}
\proof
$\bFPPqb$ is a fair quorum system since both its components are
fair, so by Proposition~\ref{pFair} we have 
\begin{eqnarray*}
 \LL{\bFPPqb} &=& \frac{\cc{\bFPPqb}}{n} \\
  &=& \frac{(3b+1)(q+1)}{(4b+1)(q^2+q+1)} \approx  \frac{3}{4q}.
\end{eqnarray*}
On the other hand, for $b$-masking systems with $n\approx 4bq^2$
elements the lower bound of Theorem~\ref{thmLoadBound} gives
\[
 \LL{\bFPPqb} \ge \sqrt{\frac{2b}{n}} \approx \frac{1}{\sqrt{2}q}.
 \inQED
\]

Note that the optimality of the load holds for {\em any} choice of $q$
and~$b$. Therefore when the number of servers (or elements) increases,
the \bFPPqb\ system can scale up using different policies while
maintaining load optimality. There are two
extremal policies: 
\begin{enumerate}
\item Fix $q$ and increase $b$; then the system can mask more failures
  when new servers are added, however the load on the servers does not
  decrease.
\item Fix $b$ and increase $q$; then the load decreases 
  when new servers are added, but the number of failures that the
  system can mask remains unchanged.
\end{enumerate}

It is important to note that systems of arbitrarily high resilience
can be constructed using the first policy since $b$ can be chosen
independently of $q$.  In particular, we can choose $b=q^a$ for any
$a>0$. Then the resulting system has $n\approx 4bq^2 =
4b^{\frac{a+2}{a}}$, and so $b \approx
\left(\frac{n}{4}\right)^{\frac{a}{a+2}}$, thus asymptotically
approaching the resilience upper bound of $\frac{n}{4}$.

Finally we analyze the crash probability of \bFPP. The following
proposition shows that \bFPP\ has good availability as long as
$p<1/4$.

\begin{proposition}
  If $p<1/4$ then $\FP{\bFPPqb} \le \exp(-\Omega(b-\log q))$.
\end{proposition}
\proof
We start by estimating $\FP{\Thresh}$.
Let \crashed\ denote the number of crashed elements in a universe of
size $4b+1$. Let $\gamma=\frac{b+1}{4b+1}-p$, thus $0<\gamma<1$ when
$p<1/4$. Then using the Chernoff bound we obtain
\begin{eqnarray} 
 \FP{\Thresh} &=& \Pr(\crashed\ge b+1) \nonumber \\
   &=&  \Pr(\crashed\ge(p+\gamma)(4b+1)) \nonumber \\
  &\le& e^{-2(4b+1)\gamma^2} \approx  e^{-b(1-4p)^2/2}.
    \label{e2} 
\end{eqnarray}
Next we estimate $\FP{\FPP}$. Let $Q_0\in\FPP$ be some quorum. Then
\begin{eqnarray}
  &&\FP{FPP} = 1-\Pr(\exists Q\in\FPP: Q{\rm\ is\ alive}) \le \nonumber\\
  &&\hspace*{-.2in}
      1-\Pr(Q_0 {\rm\ is\ alive}) = 1-(1-p)^{q+1} \le (q+1)p.
  \label{e3} 
\end{eqnarray}
Using Theorem~\ref{thmComposition} we plug (\ref{e2}) into (\ref{e3}) 
to obtain
\[
\FP{\bFPPqb} \le (q+1)e^{-b(1-4p)^2/2} = e^{-\Omega(b-\log q)}.
\inQED
\]
\Remarks
\begin{itemize}
\item
In general the crash probability is not optimal; 
since $\MT{\bFPPqb}\approx bq$ then the lower bound of
Proposition~\ref{pMT} shows we could hope for a crash probability of
$\exp(-\Omega(bq))$. Nevertheless if $q$ is constant then
\Fp\ is asymptotically optimal, and if $b\gg q$ then the gap between the
upper and lower bounds is small.
\item The final estimate we get for \FP{\bFPP}\ seems poor, as the
bound is {\em higher} than the crash probability of the
\Thresh\ components.  However this is not an artifact of
over-estimates in our analysis. Rather, it is a result of the property
that the crash probability of \FPP\ is higher than $p$, and in fact
$\FP{\FPP}\to 1$ as shown by \cite{rst92,woo96}. In this light it is
not surprising that \bFPP\ does not have an optimal crash probability.
\item The requirement $p<1/4$ is essential for this system; if $p>1/4$
  then in fact $\FP{\bFPP}$ $\to 1$ as $n\to\infty$.
\end{itemize}

\section{The multi-path system} \label{mpaths} 

Here we introduce the construction we call the Multi-Path system,
denoted by \Mpath. The elements of this system are the vertices of a
triangulated square $\sqn\times\sqn$ grid; formally, the vertices are
the points $\{(i,j)\in \RR^2: 1\le i,j\le \sqn;~ i,j\in\ZZ\}$. 
The triangulated grid has an edge between $(i_1,j_1)$ and $(i_2,j_2)$
if one of the following three conditions holds: 
(i)   $i_1=i_2$ and $j_2 = j_1+1$; 
(ii)  $j_1=j_2$ and $i_2 = i_1+1$; 
(iii) $i_2=i_1-1$ and $j_2=j_1+1$.
A quorum in the \Mpath\ system consists of
$\sqrt{2b+1}$ disjoint paths from the left side to the right side of
the grid (LR paths) and $\sqrt{2b+1}$ disjoint top-bottom (TB) paths
(see Figure~\ref{figMpath}).

\filefig
 {A multi-path construction on a $9\times 9$ grid, $b=4$, with one
   quorum shaded.}
 {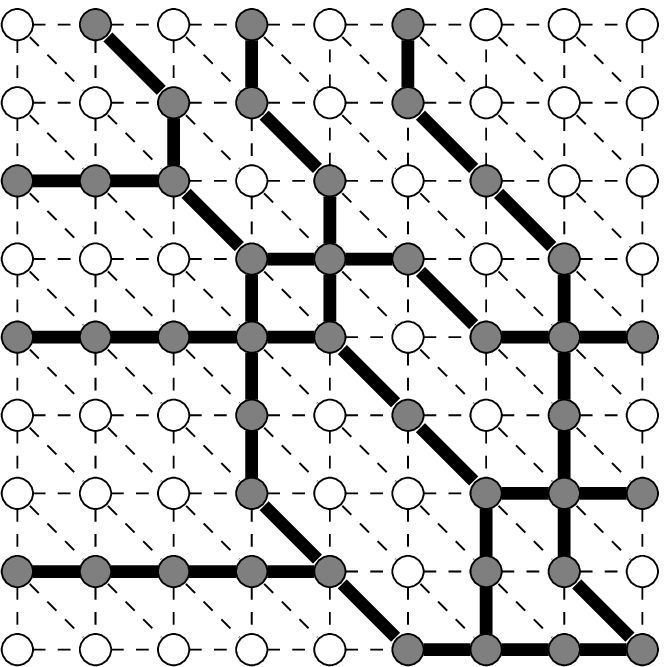}
 {figMpath}
 {2.5in} 

The \Mpath\ system has several characteristics similar to the basic
\Mgrid\ system of Section~\ref{simple}, namely an ability to mask
$b=O(\sqn)$ failures, and optimal load. Its major advantage is that
it also has an optimal crash probability \Fp. Moreover, it is the only
construction we have for which $\Fp\to 0$ as $n\to\infty$ when the
individual crash probability $p$ is arbitrarily close to 1/2. We are
able to prove this behavior of \Fp\ using results from Percolation
theory \cite{kes82,gri89}.

\Remark The system we present here is based on a triangular lattice,
with elements corresponding to vertices, as in \cite{wb92,baz96}. We
have also constructed a second system which is based on the square lattice
with elements corresponding to the edges, as in \cite{nw98}. The
properties of this second system are almost identical to those of
\Mpath, so we omit it.

\begin{proposition}
  \Mpathb\ has
  minimal quorums of size $\cc{\Mpath}\le 2\sqrt{n(2b+1)}$,
  minimal intersections of size $\IS{\Mpath} \ge 2b+1$, and
  minimal transversals of size $\MT{\Mpath}$ $=$ $\sqn-\sqrt{2b+1}+1$.
  Therefore \Mpath\ is a $b$-masking quorum system for
  $b\le\sqn-\sqrt{2}n^{1/4}$.
\end{proposition}
\proof
Let $Q_1, Q_2 \in\Mpathb$. Then the $\sqtwo$ LR paths of $Q_1$
intersect the $\sqtwo$ TB paths of $Q_2$ in $\ge 2b+1$ elements, since
the LR and TB paths are disjoint.  As in the \Mgrid\ system we have
that $\MT{\Mpathb}=\sqn-\sqtwo+1$, so when $b\le\sqn-\sqrt{2}n^{1/4}$
it follows that $\MT{\Mpathb}\ge b+1$ and we are done.
\QED

\begin{proposition}
$\LL{\Mpathb} \le 2\sqrt{\frac{2b+1}{n}}$, 
which is optimal.
\end{proposition}
\proof
The strategy only uses straight line LR and TB paths. It picks \sqtwo\
of the \sqn\ rows uniformly at random and likewise for the
columns. Clearly the load equals
the probability of accessing some element in position $i,j$, which is
\begin{eqnarray*}
 \LL{\Mpath} &\le& \Pr({\rm row\ }i{\rm\ chosen})+
    \Pr({\rm column\ }j{\rm\ chosen}) \\
 &\le&  2{{\sqn-1}\choose{\sqtwo-1}}/{\sqn\choose \sqtwo}  \\
 &=& 2\sqtwo/\sqn.
\end{eqnarray*}
By Corollary~\ref{corLowBound} this is optimal.
\QED

\begin{proposition}
 $\FP{\Mpathb} \le \exp(-\Omega(\sqn-\sqrt{b}))$ for any $p<1/2$, which is
 optimal for systems with resilience $f=O(\sqn-\sqrt{b})$.
\end{proposition}
\proof
We use the notation $\Pr_p(\cE)$ to denote the probability of event
$\cE$ defined on the grid when the individual crash probability is $p$.
A path is called ``open'' if all its elements are alive.

Let $LR$ be the event ``there exists an open LR path in the grid'', and let 
$LR_k$ be the event ``there exist $k$ open LR paths''. A failure
configuration in \Mpathb\ is one in which either $\sqtwo$ open 
LR paths or $\sqtwo$ open TB paths do not exist. By symmetry we have that
\begin{equation}
  \label{e71}
  \FP{\Mpathb} \le 2\Pr_p(\overline{LR_{\sqtwo}}) = 2(1-\Pr_p(LR_{\sqtwo})).
\end{equation}
Fix some $p'$ such that $p<p'<1/2$. Then by Theorem~\ref{tPercInc}
(see Appendix) we have that
\begin{equation}
  \label{e72}
   1-\Pr_p(LR_{\sqtwo}) \le 
    \left(\frac{1-p}{p'-p}\right)^{\sqtwo-1}[1-\Pr_{p'}(LR)].
\end{equation}
Plugging the bound on $\Pr_{p'}(LR)$ from Theorem~\ref{tPrLR} into
(\ref{e72}) and (\ref{e71}) yields 
\begin{eqnarray*}
\FP{\Mpathb} &\le& 
  2\left(\frac{1-p}{p'-p}\right)^{\sqtwo-1}e^{-\psi(p')\sqn} \\
  &=& 2e^{-\psi(p')\sqn+(\sqtwo-1)\ln{\left(\frac{1-p}{p'-p}\right)}}
\end{eqnarray*}
for some function $\psi(p')>0$.
Now $\sqtwo=O(n^{1/4})$, so for large enough $n$ we can certainly
write 
\[
\FP{\Mpathb} \le \exp(-\Omega(\sqn-\sqrt{b})).
\] 
This is optimal by Proposition~\ref{pMT}.
\QED

\section{Discussion}  \label{discussion}

We have presented four novel constructions of $b$-masking quorum
systems.  For the first time in this context, we considered the
resilience of such systems to crash failures in addition to their
tolerance of (possibly fewer) Byzantine failures.  Each of our
constructions is optimal in either its load or its crash probability
(for sufficiently small $p$).  Moreover, one of our constructions,
namely M-Paths, is optimal in both measures.  One of our constructions
is achieved using a novel boosting technique that makes all known
benign fault-tolerant quorum constructions available for Byzantine
environments (of appropriate sizes).  In proving optimality of our
constructions, we also contribute lower bounds on the load and crash
probability of any $b$-masking quorum system.

\begin{table}
\begin{center}
\begin{tabular}{|l|c|c|c|c|}
\hline
\multicolumn{1}{|c|}{System} & $b <$ & $f$ & $\L$ & $F_p$ \\ \hline
Threshold~\cite{mr98a}   & $n/4$
                        & $O(n-b)$
                        & $1/2 + O(b/n)$
                        & $\exp(-\Omega(f))$ {\footnotesize $^{\ast}$}
                        \\ \hline
Grid~\cite{mr98a}        & $\sqrt{n}/3$
                        & $O(\sqrt{n} - b)$
                        & $O(b/\sqrt{n})$
                        & $\tends{n} 1$
                        \\ \hline \hline
\Mgrid\                 & $\sqrt{n}/2$
                        & $O(\sqrt{n} - \sqrt{b})$
                        & $O(\sqrt{b/n})$  {\footnotesize $^{\dag}$}
                        & $\tends{n} 1$
                        \\ \hline
\RT($k$, $\ell$){\footnotesize $^{\ddag}$}
                        & $O(\min\{n^{\alpha_1},n^{\alpha_2}\})$
                                 {\footnotesize $^{\ddag}$}
                        & $O(b)$
                        & $n^{-(1-\log_k\ell)}$ 
                        & $\exp(-\Omega(f))$ {\footnotesize $^{\ast}$}
                        \\ \hline
\bFPP\                  & $n/4$
                        & $O(\sqrt{bn})$
                        & $O(\sqrt{b/n})$  {\footnotesize $^{\dag}$}
                        & $\exp(-\Omega(b-\log(n/b)))$
                        \\ \hline
\Mpath\                 & $(1-o(1))\sqn$
                        & $O(\sqrt{n} - \sqrt{b})$
                        & $O(\sqrt{b/n})$  {\footnotesize $^{\dag}$}
                        & $\exp(-\Omega(f))$ {\footnotesize $^{\ast}$}
                        \\ \hline
\multicolumn{4}{l}{\footnotesize $^{\dag}$ Optimal for $b$-masking systems} \\
\multicolumn{4}{l}{\footnotesize $^{\ast}$ Optimal for $f$-resilient systems}\\
\multicolumn{4}{l}{\footnotesize $^{\ddag}$ $\alpha_1=\log_k(2\ell - k)$ and
$\alpha_2=\log_k(k-\ell+1)$}
\end{tabular}
\end{center}
\caption{Constructions in this paper ($n$ = number of servers).}
\label{results}
\end{table}

The properties of our various constructions are summarized in
Table~\ref{results}, alongside the properties of two other
$b$-masking constructions proposed in~\cite{mr98a}, namely Threshold
and Grid. 

Determining the best quorum construction depends on the goals and
constraints of any particular settings, as no system is advantageous
in all measures. For example, suppose we fix $n$ to be $1024$, the
desired load $\L$ to be approximately $1/4$, and assume that the
individual failure probability of components is $1/8$.  
In these
settings, an \Mgrid\ system can tolerate $b=15$ Byzantine failures and
up to $f=28$ benign failures, but has a failure probability $F_p \ge
0.638$. 
In the same
settings, a \bFPP\ system (with $n=1001$, $q=3$) can tolerate $b=19$,
up to $f=79$ benign failures, with somewhat better failure
probability: it has $F_p \le 0.372$. 
The \Mpath\ construction, with 4~LR and 4~TB paths per quorum, has
$b=7$ here, and can tolerate up to $f=29$ benign failures, but has a
good crash probability: $F_p \le 0.001$ (using the estimate following
Theorem~\ref{tPrLR}, together with Theorem~\ref{tPercInc} with
$p'=1/7$).
In this setting, the $\RT(4,3)$ construction, with depth $h=5$, is the
best, with $b=15$, $f=31$ and an excellent failure probability of only
$F_p \le 0.0001$.

More generally, if masking large numbers of Byzantine server failures
is important, then of the systems listed in Table~\ref{results}, only
Threshold and \bFPP\ can provide the highest possible masking
ability, i.e., up to $b < n/4$.  However, Threshold can mask $n/4$
Byzantine failures for any system size, whereas \bFPP\ approaches such
degree of Byzantine resilience only for very large $n$. If, on the
other hand, load is more crucial, then Threshold suffers in load
whereas \bFPP\ offers reduced load, as do the other three systems in
this paper, albeit with lower masking ability.  If masking
fewer Byzantine server failures is allowable, then other quorum
constructions can be used, in particular \RT\ and \Mpath.  These two
constructions have similar masking ability, resilience, and load, but
\Mpath\ has asymptotically superior crash probability when $p$ is
close to 1/2.

Finally, we note that it is impossible to achieve optimal resilience
and load simultaneously: Since necessarily $f \le \ccQ$,
Theorem~\ref{thmLoadBound} implies that $f \le n\LQ$, i.e., when load
is low then so is resilience, and when resilience is high then so is
load.  In order to break this tradeoff, in~\cite{mrww98} we propose
relaxing the intersection property of masking quorum systems, so that
``quorums'' chosen according to a specific strategy intersect each
other in
enough correct servers to maintain correctness of the system with a high
probability.

\section*{Acknowledgments}
We thank Oded Goldreich and the anonymous referees of the 16th ACM
Symposium on Principles of Distributed Computing for many helpful
comments on an earlier version of this paper.

{ \small
\newcommand{\etalchar}[1]{$^{#1}$}
\newcommand{\nopsort}[1]{}
\newcommand{\lb}{\linebreak[0]}

}


\appendix
\centerline{\LARGE\bf Appendix}

\section{Combinatorial Lemmas}

\begin{lemma} \label{lApp0} 
  Let $0\le i,d \le k$ be integers. Then
   $\frac{{k\choose d+i}}{{k\choose d}} \le {{k-d}\choose i}$.
\end{lemma}
\proof
\[
\frac{{k\choose d+i}}{{k\choose d}} =
  \frac{k!d!(k-d)!}{(d+i)!(k-d-i)!k!} = 
    \frac{(k-d)!}{(k-d-i)!}\frac{d!}{(d+i)!}
 \le \frac{(k-d)!}{(k-d-i)!i!} = {{k-d}\choose i}
\inQED
\]

\begin{lemma} \label{lApp1} 
Let $0\le d \le k$ be integers and let $p\in[0,1]$. Then
\[
  \sum_{j=d}^k {k\choose j} p^j(1-p)^{k-j} \le {k\choose d}p^d.
\]
\end{lemma}
\proof
\[
 \sum_{j=d}^k {k\choose j} p^j(1-p)^{k-j} = 
  {k\choose d}p^d \sum_{j=d}^k \frac{{k\choose j}}{{k\choose d}} 
         p^{j-d}(1-p)^{k-j},
\]
so it suffices to show that the last sum is $\le 1$. But using
Lemma~\ref{lApp0} we get
\begin{eqnarray*}
 && \sum_{j=d}^k \frac{{k\choose j}}{{k\choose d}}p^{j-d}(1-p)^{k-j} =
  \sum_{i=0}^{k-d}\frac{{k\choose d+i}}{{k\choose d}}p^i(1-p)^{k-d-i}
  \\
 &\le& \sum_{i=0}^{k-d} {{k-d}\choose i} p^i(1-p)^{k-d-i} =
  [p+(1-p)]^{k-d}=1.
\inQED
\end{eqnarray*}

\section{Theorems of Percolation Theory}

In this section we list the definitions and results that are used in
our analysis of the \Mpath\ system, following \cite{kes82,gri89}.

The percolation model we are interested in is as follows. Let $\ZZ$ be
the graph of the (infinite) triangle lattice in the plane. Assume that
a vertex is {\em closed\/} with probability $p$ and {\em open\/} with
probability $1-p$, independently of other vertices. This model is
known as site percolation on the triangle lattice. Another natural
model, which plays a minor role in our work, is the bond percolation
model. In it the {\em edges} are closed with probability $p$.

A key idea in percolation theory is that there exists a {\em critical
probability}, $p_c$, such that graphs with $p<p_c$ exhibit
qualitatively different properties than graphs with $p>p_c$. For
example, $\ZZ$ with $p<p_c$ has a single connected (open) component of
infinite size. When $p>p_c$ there is no such component.  For site
percolation on the triangle $p_c = 1/2$ \cite{kes80}.

The following theorem shows that when the probability $p$ 
for a closed vertex is below the critical probability, the probability of
having long open paths tends to~1 exponentially fast. Recall that 
$LR$ is the event ``there exists an open LR path in the
$\sqn\times\sqn$ grid''. Then \cite{men86} (see also \cite{gri89}
p.~287) implies  

\begin{theorem} \label{tPrLR} 
  If $p<1/2$ then $\Pr_p(LR) \ge 1-e^{-\psi(p)\sqn}$, for some $\psi(p)>0$
  independent of $n$.
\end{theorem}

\Remark The dependence of $\psi$ on $p$ is such that $\psi(p)\to 0$
when $p\to 1/2$. However, for $p$'s not too close to $1/2$ we can obtain
concrete estimates using elementary techniques. For instance,
a counting argument similar to that of Bazzi \cite{baz96} shows that
\[
\Pr_p(LR) \ge 1-\frac{\sqn(3p)^\sqn}{1-3p},
\]
when $p < 1/3$.

\begin{definition}
Let $\cE$ be an event defined in the percolation model. Then the 
{\em interior} of $\cE$ with depth $r$,
denoted $I_r(\cE)$, is the set of all configurations
in $\cE$ which are still in $\cE$ even if we perturb the states of up
to $r$ vertices.
\end{definition}

We may think of $I_r(\cE)$ as the event that $\cE$ occurs and is
`stable' with respect to changes in the states of $r$ or fewer vertices.
The definition is useful to us in the following situation.
If $LR$ is the event ``there exists an open left-right path in a
rectangle~$D$'', then it follows that 
$I_r(LR)$ is the event ``there are at least $r+1$ disjoint open 
left-right paths in $D$''.

\begin{theorem} \label{tPercInc} 
{\rm \cite{accfr83}}
Let $\cE$ be an increasing event and let $r$ be a positive integer.
Then 
\[
 1-\Pr_p(I_r(\cE)) \le \left(\frac{1-p}{p'-p}\right)^r[1-\Pr_{p'}(\cE)]
\]
whenever $0 \le p < p' \le 1$.
\end{theorem}

The theorem amounts to the assertion that if $\cE$ is likely to occur
when the crash probability is $p'$, then $I_r(\cE)$ is likely to
occur when the crash probability $p$ is smaller than $p'$.

\end{document}

%% file: grid.tex
\setlength{\unitlength}{0.0125in}
\begingroup\makeatletter\ifx\SetFigFont\undefined
\def\x#1#2#3#4#5#6#7\relax{\def\x{#1#2#3#4#5#6}}%
\expandafter\x\fmtname xxxxxx\relax \def\y{splain}%
\ifx\x\y   
\gdef\SetFigFont#1#2#3{%
  \ifnum #1<17\tiny\else \ifnum #1<20\small\else
  \ifnum #1<24\normalsize\else \ifnum #1<29\large\else
  \ifnum #1<34\Large\else \ifnum #1<41\LARGE\else
     \huge\fi\fi\fi\fi\fi\fi
  \csname #3\endcsname}%
\else
\gdef\SetFigFont#1#2#3{\begingroup
  \count@#1\relax \ifnum 25<\count@\count@25\fi
  \def\x{\endgroup\@setsize\SetFigFont{#2pt}}%
  \expandafter\x
    \csname \romannumeral\the\count@ pt\expandafter\endcsname
    \csname @\romannumeral\the\count@ pt\endcsname
  \csname #3\endcsname}%
\fi
\fi\endgroup
\begin{picture}(140,155)(0,-10)
\path(140,80)(140,100)(120,100)
	(120,80)(140,80)
\path(40,100)(40,120)(20,120)
	(20,100)(40,100)
\path(60,100)(60,120)(40,120)
	(40,100)(60,100)
\path(120,80)(120,100)(100,100)
	(100,80)(120,80)
\path(20,60)(20,80)(0,80)
	(0,60)(20,60)
\path(40,60)(40,80)(20,80)
	(20,60)(40,60)
\path(80,60)(80,80)(60,80)
	(60,60)(80,60)
\path(60,0)(60,20)(40,20)
	(40,0)(60,0)
\path(140,40)(120,40)(120,60)
	(140,60)(140,40)
\path(20,80)(0,80)(0,100)
	(20,100)(20,80)
\path(60,80)(40,80)(40,100)
	(60,100)(60,80)
\path(40,80)(20,80)(20,100)
	(40,100)(40,80)
\path(80,80)(60,80)(60,100)
	(80,100)(80,80)
\path(60,60)(40,60)(40,80)
	(60,80)(60,60)
\path(20,40)(0,40)(0,60)
	(20,60)(20,40)
\path(40,40)(20,40)(20,60)
	(40,60)(40,40)
\path(80,40)(60,40)(60,60)
	(80,60)(80,40)
\path(120,40)(100,40)(100,60)
	(120,60)(120,40)
\path(20,0)(0,0)(0,20)
	(20,20)(20,0)
\path(20,20)(0,20)(0,40)
	(20,40)(20,20)
\path(80,20)(60,20)(60,40)
	(80,40)(80,20)
\path(40,20)(20,20)(20,40)
	(40,40)(40,20)
\path(60,40)(40,40)(40,60)
	(60,60)(60,40)
\path(40,0)(40,20)(20,20)
	(20,0)(40,0)
\path(60,20)(40,20)(40,40)
	(60,40)(60,20)
\path(80,0)(80,20)(60,20)
	(60,0)(80,0)
\texture{cccccccc 0 0 0 cccccccc 0 0 0 
	cccccccc 0 0 0 cccccccc 0 0 0 
	cccccccc 0 0 0 cccccccc 0 0 0 
	cccccccc 0 0 0 cccccccc 0 0 0 }
\shade\path(100,80)(80,80)(80,100)
	(100,100)(100,80)
\path(100,80)(80,80)(80,100)
	(100,100)(100,80)
\shade\path(100,40)(80,40)(80,60)
	(100,60)(100,40)
\path(100,40)(80,40)(80,60)
	(100,60)(100,40)
\shade\path(100,0)(80,0)(80,20)
	(100,20)(100,0)
\path(100,0)(80,0)(80,20)
	(100,20)(100,0)
\shade\path(100,100)(80,100)(80,120)
	(100,120)(100,100)
\path(100,100)(80,100)(80,120)
	(100,120)(100,100)
\shade\path(100,60)(80,60)(80,80)
	(100,80)(100,60)
\path(100,60)(80,60)(80,80)
	(100,80)(100,60)
\shade\path(100,20)(80,20)(80,40)
	(100,40)(100,20)
\path(100,20)(80,20)(80,40)
	(100,40)(100,20)
\path(0,120)(0,100)(0,100)
	(0,120)(0,120)
\path(120,20)(100,20)(100,40)
	(120,40)(120,20)
\shade\path(40,20)(20,20)(20,40)
	(40,40)(40,20)
\path(40,20)(20,20)(20,40)
	(40,40)(40,20)
\path(120,60)(120,80)(100,80)
	(100,60)(120,60)
\shade\path(20,0)(0,0)(0,20)
	(20,20)(20,0)
\path(20,0)(0,0)(0,20)
	(20,20)(20,0)
\shade\path(20,100)(0,100)(0,120)
	(20,120)(20,100)
\path(20,100)(0,100)(0,120)
	(20,120)(20,100)
\shade\path(20,40)(0,40)(0,60)
	(20,60)(20,40)
\path(20,40)(0,40)(0,60)
	(20,60)(20,40)
\shade\path(20,80)(0,80)(0,100)
	(20,100)(20,80)
\path(20,80)(0,80)(0,100)
	(20,100)(20,80)
\path(120,0)(120,20)(100,20)
	(100,0)(120,0)
\path(140,0)(140,20)(120,20)
	(120,0)(140,0)
\path(140,20)(140,40)(120,40)
	(120,20)(140,20)
\path(140,100)(140,120)(120,120)
	(120,100)(140,100)
\path(140,60)(140,80)(120,80)
	(120,60)(140,60)
\path(80,100)(80,120)(60,120)
	(60,100)(80,100)
\path(120,100)(120,120)(100,120)
	(100,100)(120,100)
\path(140,120)(140,140)(120,140)
	(120,120)(140,120)
\path(120,120)(120,140)(100,140)
	(100,120)(120,120)
\path(80,120)(80,140)(60,140)
	(60,120)(80,120)
\path(60,120)(60,140)(40,140)
	(40,120)(60,120)
\path(40,120)(40,140)(20,140)
	(20,120)(40,120)
\path(20,120)(20,140)(0,140)
	(0,120)(20,120)
\shade\path(100,120)(80,120)(80,140)
	(100,140)(100,120)
\path(100,120)(80,120)(80,140)
	(100,140)(100,120)
\shade\path(20,120)(0,120)(0,140)
	(20,140)(20,120)
\path(20,120)(0,120)(0,140)
	(20,140)(20,120)
\shade\path(140,60)(120,60)(120,80)
	(140,80)(140,60)
\path(140,60)(120,60)(120,80)
	(140,80)(140,60)
\shade\path(120,60)(100,60)(100,80)
	(120,80)(120,60)
\path(120,60)(100,60)(100,80)
	(120,80)(120,60)
\shade\path(80,60)(60,60)(60,80)
	(80,80)(80,60)
\path(80,60)(60,60)(60,80)
	(80,80)(80,60)
\shade\path(60,60)(40,60)(40,80)
	(60,80)(60,60)
\path(60,60)(40,60)(40,80)
	(60,80)(60,60)
\shade\path(40,60)(20,60)(20,80)
	(40,80)(40,60)
\path(40,60)(20,60)(20,80)
	(40,80)(40,60)
\shade\path(20,60)(0,60)(0,80)
	(20,80)(20,60)
\path(20,60)(0,60)(0,80)
	(20,80)(20,60)
\shade\path(20,20)(0,20)(0,40)
	(20,40)(20,20)
\path(20,20)(0,20)(0,40)
	(20,40)(20,20)
\shade\path(80,20)(60,20)(60,40)
	(80,40)(80,20)
\path(80,20)(60,20)(60,40)
	(80,40)(80,20)
\shade\path(120,20)(100,20)(100,40)
	(120,40)(120,20)
\path(120,20)(100,20)(100,40)
	(120,40)(120,20)
\shade\path(140,20)(120,20)(120,40)
	(140,40)(140,20)
\path(140,20)(120,20)(120,40)
	(140,40)(140,20)
\shade\path(60,20)(40,20)(40,40)
	(60,40)(60,20)
\path(60,20)(40,20)(40,40)
	(60,40)(60,20)
\end{picture}

%% file: lofk.tex
\setlength{\unitlength}{0.0125in}
\begingroup\makeatletter\ifx\SetFigFont\undefined
\def\x#1#2#3#4#5#6#7\relax{\def\x{#1#2#3#4#5#6}}%
\expandafter\x\fmtname xxxxxx\relax \def\y{splain}%
\ifx\x\y   
\gdef\SetFigFont#1#2#3{%
  \ifnum #1<17\tiny\else \ifnum #1<20\small\else
  \ifnum #1<24\normalsize\else \ifnum #1<29\large\else
  \ifnum #1<34\Large\else \ifnum #1<41\LARGE\else
     \huge\fi\fi\fi\fi\fi\fi
  \csname #3\endcsname}%
\else
\gdef\SetFigFont#1#2#3{\begingroup
  \count@#1\relax \ifnum 25<\count@\count@25\fi
  \def\x{\endgroup\@setsize\SetFigFont{#2pt}}%
  \expandafter\x
    \csname \romannumeral\the\count@ pt\expandafter\endcsname
    \csname @\romannumeral\the\count@ pt\endcsname
  \csname #3\endcsname}%
\fi
\fi\endgroup
\begin{picture}(265,92)(0,-10)
\put(75,5){\ellipse{10}{10}}
\put(145,5){\ellipse{10}{10}}
\put(160,5){\ellipse{10}{10}}
\put(175,5){\ellipse{10}{10}}
\put(190,5){\ellipse{10}{10}}
\put(260,5){\ellipse{10}{10}}
\texture{cccccccc 0 0 0 cccccccc 0 0 0 
	cccccccc 0 0 0 cccccccc 0 0 0 
	cccccccc 0 0 0 cccccccc 0 0 0 
	cccccccc 0 0 0 cccccccc 0 0 0 }
\put(5,5){\shade\ellipse{10}{10}}
\put(5,5){\ellipse{10}{10}}
\put(35,5){\shade\ellipse{10}{10}}
\put(35,5){\ellipse{10}{10}}
\put(50,5){\shade\ellipse{10}{10}}
\put(50,5){\ellipse{10}{10}}
\put(90,5){\shade\ellipse{10}{10}}
\put(90,5){\ellipse{10}{10}}
\put(105,5){\shade\ellipse{10}{10}}
\put(105,5){\ellipse{10}{10}}
\put(120,5){\shade\ellipse{10}{10}}
\put(120,5){\ellipse{10}{10}}
\put(215,5){\shade\ellipse{10}{10}}
\put(215,5){\ellipse{10}{10}}
\put(230,5){\shade\ellipse{10}{10}}
\put(230,5){\ellipse{10}{10}}
\put(245,5){\shade\ellipse{10}{10}}
\put(245,5){\ellipse{10}{10}}
\path(130,60)(100,40)
\path(135,60)(165,40)
\path(130,60)(30,40)
\path(135,60)(235,40)
\path(25,25)(20,10)
\path(25,25)(5,10)
\path(30,25)(35,10)
\put(20,5){\ellipse{10}{10}}
\path(30,25)(50,10)
\put(119,64){\makebox(0,0)[lb]{\smash{{{\SetFigFont{10}{12.0}{rm}3 of 4}}}}}
\path(115,25)(115,40)(80,40)
	(80,25)(115,25)
\path(95,25)(90,10)
\path(95,25)(75,10)
\path(100,25)(105,10)
\path(100,25)(120,10)
\path(45,25)(45,40)(10,40)
	(10,25)(45,25)
\path(185,25)(185,40)(150,40)
	(150,25)(185,25)
\path(165,25)(160,10)
\path(165,25)(145,10)
\path(170,25)(175,10)
\path(170,25)(190,10)
\path(255,25)(255,40)(220,40)
	(220,25)(255,25)
\path(235,25)(230,10)
\path(235,25)(215,10)
\path(240,25)(245,10)
\path(240,25)(260,10)
\path(150,60)(150,75)(115,75)
	(115,60)(150,60)
\put(84,29){\makebox(0,0)[lb]{\smash{{{\SetFigFont{10}{12.0}{rm}3 of 4}}}}}
\put(14,29){\makebox(0,0)[lb]{\smash{{{\SetFigFont{10}{12.0}{rm}3 of 4}}}}}
\put(154,29){\makebox(0,0)[lb]{\smash{{{\SetFigFont{10}{12.0}{rm}3 of 4}}}}}
\put(224,29){\makebox(0,0)[lb]{\smash{{{\SetFigFont{10}{12.0}{rm}3 of 4}}}}}
\end{picture}